\documentstyle[preprint,aps,pra,psfig]{revtex}

\begin{document}
\tighten
\draft

\title{Collisional Frequency Shifts of Absorption Lines in an Atomic
       Hydrogen Gas}
\author{C.\ J.\ Pethick$^1$ and H.\ T.\ C.\ Stoof$^2$}

\address{$^1$ NORDITA, Blegdamsvej 17, DK-2100 Copenhagen \O, Denmark, \\
         $^2$ Institute for Theoretical Physics, University of Utrecht, \\
              Princetonplein 5, 3584 CC  Utrecht, The Netherlands}

\maketitle

\begin{abstract}
We consider the effect of interactions on the line shape of the
two-photon  $1s-2s$ transition in a (doubly) spin-polarized atomic hydrogen
gas in terms of the interatomic interaction potentials. We show that the
frequency-weighted sum rule for the intensity of the line is not given
simply in terms of the pseudopotentials that describe the interactions
between low-energy atoms. The origin of the departures from the simple
pseudopotential result for the frequency-weighted sum rule is traced to
what we refer to as incoherent contributions to the spectral
weight. These arise from more complicated final states of
the many-body system than the ones usually considered. In particular, we show
how the relevant response function may be treated in a manner similar to the
density-density response function for Fermi liquids, and express it as a
coherent part coming from single particle-hole pairs, and an incoherent part
coming from other excitations. We argue that in experiments only the coherent
part of the response of the system is observed, and its contribution to the
frequency-weighted sum rule is shown to be given correctly by the
pseudopotential approximation. Finally we calculate the width of the coherent
part of the line due to collisional damping. \\
\end{abstract}


\section{Introduction}
\label{int}
After two decades of concentrated effort Fried {\it et al.} recently succeeded
in realizing Bose-Einstein condensation in spin-polarized atomic hydrogen in a
magnetic trap \cite{mit1}. In this experiment a key role is played by
collisional frequency shifts, since the density of the atomic hydrogen cloud is
monitored by observing the frequency shift of the Doppler-free peak in the
two-photon $1s-2s$ absorption spectrum \cite{mit2}. Previously, collisional
frequency shifts have also been observed in hydrogen masers \cite{masers} and in
atomic fountains \cite{fountains}, where they lead to a serious limitation on
the stability of these devices. A thorough understanding of such shifts is
therefore central to the interpretation of various experimental results with
atomic quantum gases.

In the theory of line shifts that is currently standard \cite{standard}, one
considers only the normal state of the gas and uses a Boltzmann equation to
determine the effect of collisions on the absorption profile. The line shift
is then found to be proportional to the difference of the $1s-2s$ scattering
length, $a_{1s-2s}$, and the $1s-1s$ one, $a_{1s-1s}$.
Recently two papers have appeared on the theory of the line shifts, one
employing the random phase approximation \cite{ol}, and the other using
sum-rule arguments \cite{levitov}. One striking prediction of these
calculations is that the line shift in a dilute, fully Bose-Einstein condensed
gas should be
one half that for an uncondensed gas of the same
density. These papers have in common the assumption that the interaction
between atoms may be assumed to be of the usual contact pseudopotential form,
and that the interactions may be taken into account in a mean-field approach.
In this paper we investigate the problem allowing for a more general
interaction. We demonstrate that the frequency-weighted sum rule is given in
terms of the {\em bare} interaction potential, not the pseudopotential. By using
microscopic many-body theory we trace the origin of the discrepancy
between the true frequency-weighted sum rule and the one calculated using the
pseudopotential to {\em incoherent} contributions to the atomic propagators,
which arise when a 1$s$ atom is excited close to
another such atom. The latter processes, while relatively infrequent in a
low-density gas, give contributions to the spectral weight at frequencies very
different from those for excitation of an atom far away
from any other atom. However, as we demonstrate in this paper, the
shift of the {\em coherent} contribution to the response, which corresponds
physically to excitation of an atom when it is relatively far away from other
atoms, is given by the pseudopotential result.

We have organized the paper as follows. In Sec.~\ref{sr} we first derive an
exact sum rule for the frequency-weighted spectral weight,
and will show that this is not given correctly by the pseudopotential result.
In Sec.~\ref{ma} we then study the problem from a microscopic point of
view, and indicate how the absorption spectrum can be separated into coherent
and
incoherent parts. We also argue that the coherent part of the response is of
greatest interest
experimentally. In Sec.~\ref{cb} we determine the collisional broadening of the
coherent absorption peak and we end in Sec.~\ref{concl} with our conclusions.

\section{Sum rule approach}
\label{sr}
Let us begin by considering a system of hydrogen atoms in the $1s$
ground state. The effect of applying the laser radiation is to excite some
hydrogen atoms to the metastable $2s$ state, which has a radiative
lifetime $1/\Gamma_{2s}$ of the order of one second. Experimentally, the
hydrogen clouds investigated are inhomogeneous, but since the length scale for
density variations is large compared with the microscopic lengths in the
problem, it is an excellent approximation to take the effects of inhomogeneity
into account in the local density approximation, and consequently in our
calculations we consider a spatially uniform system. If the radiation field is
spatially uniform, its interaction with the hydrogen gas may be represented by a
perturbing Hamiltonian
\begin{equation}
\label{pert}
H_1=\frac{\hbar\Omega}{2} \int d{\bf x}~
  \left[ e^{-i\omega t}\psi_{2s}^{\dagger}({\bf x})\psi_{1s}({\bf x})
        + e^{i\omega t}\psi_{1s}^{\dagger}({\bf x})\psi_{2s}({\bf x}) \right]~,
\end{equation}
where $\omega$ is the angular frequency of the pair of photons, the operators
$\psi_{\alpha}^{\dagger}({\bf x})$ and $\psi_{\alpha}({\bf x})$ create and
destroy atoms in the state $|\alpha\rangle$, and $\Omega$ is the effective Rabi
frequency determined by the strength of the laser field and atomic matrix
elements.

The unperturbed part $H_0$ of the Hamiltonian is given by the sum of the
intrinsic atomic energies of isolated atoms at rest, the kinetic energy
associated with the translation of atoms, and terms that take into account
interactions between atoms. To an excellent approximation the interaction energy
is given in terms of local two-body potentials dependent only on the distance
$r$ between atoms, and we denote the potential for two atoms in the
$1s$ state by $V_{1s-1s}(r)$ and that for one atom in the
$1s$ state and the other in the $2s$ state by $V_{1s-2s}(r)$.
Since we consider the case of weak excitation, we shall not need to specify the
interaction between two excited-state atoms.
In detail we thus have
\begin{eqnarray}
H_0 &=& \int d{\bf x}~
      \psi_{1s}^{\dagger}({\bf x})
           \left( -\frac{\hbar^2 \nabla^2}{2m} +
                   \epsilon_{1s} \right) \psi_{1s}({\bf x})
     + \int d{\bf x}~
      \psi_{2s}^{\dagger}({\bf x})
           \left( -\frac{\hbar^2 \nabla^2}{2m} +
                   \epsilon_{2s}  \right) \psi_{2s}({\bf x})
                                                       \nonumber \\
    &+& \frac{1}{2} \int d{\bf x}\int d{\bf x}'~
        \psi_{1s}^{\dagger}({\bf x}) \psi_{1s}^{\dagger}({\bf x}')
           V_{1s-1s}({\bf x}-{\bf x}') \psi_{1s}({\bf x}') \psi_{1s}({\bf x})
                                                       \nonumber \\
    &+& \int d{\bf x}\int d{\bf x}'~
        \psi_{1s}^{\dagger}({\bf x}) \psi_{2s}^{\dagger}({\bf x}')
           V_{1s-2s}({\bf x}-{\bf x}') \psi_{2s}({\bf x}') \psi_{1s}({\bf x})~,
\end{eqnarray}
where $m$ is the mass of an atom and $\epsilon_{\alpha}$ denotes the energy
of the atomic state $\alpha$. Note that for clarity we have in the
first instance neglected the effect of the finite lifetime of the excited atom.
In Sec.~\ref{ma}, however, we show how it can be easily incorporated into the
theory.

The net rate of transitions may now be calculated from Fermi's Golden Rule, and
is given by
\begin{equation}
\label{gr1}
I(\omega) = \frac{2\pi}{\hbar}\sum_{m,n} |\langle m|H_1|n\rangle|^2
\delta(\hbar\omega + E_n-E_m) (p_n -p_m)~.
\label{gr}
\end{equation}
Here $p_n$ is the initial probability for occurrence of the
many-body state $|n\rangle$, which is an eigenstate of the
Hamiltonian $H_0$ and therefore obeys $H_0 |n\rangle = E_n
|n\rangle$. We note that for the situations of interest in the
Bose-Einstein condensation experiments, initial states containing
2$s$ atoms play essentially no role, since the probability of 2$s$
atoms being present is very small because the energy difference between
a $2s$ atom and a ground-state one is much larger than the thermal
energy $k_BT$. From Eq.~(\ref{gr}) we thus find that the rate of
absorption of energy is
\begin{equation}
\label{gr2}
\hbar \omega I(\omega) =\frac{2\pi}{\hbar}\sum_{m,n} (E_m-E_n)
|\langle m|H_1|n\rangle|^2 \delta(\hbar \omega -E_m+E_n) p_n~,
\end{equation}
and the average frequency of the line is given by
\begin{equation}
\label{ave1}
\bar{\omega} = \frac{\int d\omega \omega I(\omega)} {\int d\omega
I(\omega)} =\frac{ \sum_{m,n} (E_m-E_n)|\langle m|H_1|n\rangle|^2 p_n }
                 {\hbar \sum_{m,n}|\langle m|H_1|n\rangle|^2 p_n }~.
\end{equation}

To evaluate the average frequency we, following the procedure adopted by
Oktel {\it et al.} \cite{levitov}, again make use of
$H_0 |n\rangle = E_n |n\rangle$, and consider the thermal average of
$H_1[H_0,H_1]$, or
equivalently the double commutator $[H_1,[H_0,H_1]]$. In contrast to Ref.
\cite{levitov}, we however do not assume that the interaction may be represented
by a pseudopotential. The average frequency is then given by
\begin{equation}
\label{ave2}
\bar{\omega} = \frac{\langle H_1[H_0,H_1] \rangle}{\hbar \langle H_1^2 \rangle
}~. \end{equation}
The physical content of this equation is that the average frequency shift
is given by the difference in energies of the expectation value of the
energy in the initial state and that in the state created by operating
with $H_1$ on the initial state. Evaluating the expectation value of the
commutator expression above directly, we find for the frequency shift relative
to its value for an isolated atom the result
\begin{equation}
\label{aveshift}
\overline{\Delta \omega} = \frac{n}{\hbar} \int d{\bf r}~
                                   [V_{1s-2s}(r)-V_{1s-1s}(r)] g_2({\bf r})~,
\end{equation}
where $n$ is the density of the gas and
\begin{equation}
\label{corr}
g_2({\bf r})=\frac{1}{n^2}
  \left\langle \psi^{\dagger}_{1s}({\bf r}) \psi^{\dagger}_{1s}({\bf 0})
               \psi_{1s}({\bf 0}) \psi_{1s}({\bf r}) \right\rangle
\end{equation}
is the pair distribution function for ground state atoms in the initial
state of the system. In arriving at this expression we have again neglected the
possibility of 2$s$ atoms being present in the initial state. This result is
simple to understand, since the operator $H_1$ converts a single ground-state
atom in the initial state into an excited state one with an amplitude that
does not depend on position. The average energy difference between the initial
state and the one created by the laser is therefore the energy required to
convert a $1s$ atom into a $2s$ one. Since the masses of the atoms in the two
states are the same, there is no contribution from the kinetic energy, and the
sole contribution, apart from the energy difference for an isolated atom, comes
from interactions. This situation should be contrasted with that of an isotopic
impurity, like a $^3$He atom in liquid $^4$He, which is just the opposite, in 
that
the masses are different, while the interaction potentials are the same. The sum
rule derived here is analogous to sum rules for spin response of condensed
matter systems, and for spin, isospin, and spin-isospin response of nuclei. In
these cases the basic origin of the shifts is terms in the interaction that are
not invariant under rotations in spin, and/or isospin space, or, in the present
problem, rotations in the pseudospin space corresponding to conversion of a
$1s$ atom into a $2s$ one.

The long-wavelength assumption is appropriate for the two-photon transition
when the two photons that are absorbed have equal and opposite momenta.
When the total momentum ${\bf q}$ of the absorbed photons is non-zero, the
perturbing Hamiltonian depends on space and we need to generalize
the sum rule to spatially varying interactions. This is 
straightforward and we find that the average frequency shift is given by adding
the recoil energy $\hbar^2{\bf q}^2/2m$ to the spatially homogeneous result for
${\bf q} = {\bf 0}$.

Let us now compare our result in Eq.~(\ref{aveshift}) with that of earlier work.
If the interaction potentials are weak, the correlation function will vary
little over the ranges of the potentials, and we may replace the pair
distribution function by its value for zero separation. We then obtain
\begin{equation}
\label{aveshiftB}
(\overline{\Delta \omega})_{\rm B} = \frac{n}{\hbar} g_2({\bf 0})
\int d{\bf r}~[V_{1s-2s}(r)-V_{1s-1s}(r)]~.
\end{equation}
This is equivalent to the result of Levitov {\it et al.} \cite{levitov}, since
for weak potentials the Born approximation may be applied, and thus the
scattering lengths $a$ are related to the interaction potentials by
$4\pi \hbar^2 a/m=\int d{\bf r}~ V(r)$.

The interaction potentials for hydrogen atoms are not weak, and the Born
approximation is not valid. Therefore it is important to explore how the
pair distribution function behaves at short distances. On length scales
larger than the range of the atomic interactions, correlations should be
well described in terms of mean fields. However, for strong potentials it
is not permissible to assume that the correlation function for small
separations varies slowly on distances of the order of the range of the
potential. Rather one expects that the many-body wave function for small
particle separations will behave as that for a pair of atoms interacting via the
$1s-1s$ interaction, since the effects of other atoms will then be negligible at
low densities. Hence, provided the energies of elementary excitations of the
system are small compared with the typical energy scale over which the two-atom
relative wave function changes significantly, it will be a good
approximation to assume that the pair distribution function scales as the
square of the scattering wave function at zero energy, i.e.,
$|\Psi_{\rm rel}(r)|^2$. We therefore write
\begin{equation}
g_2({\bf r}) \simeq  |\Psi_{\rm rel}(r)|^2 g^{\rm MF}_2({\bf 0})~,
\end{equation}
where the mean-field correlation function $g^{\rm MF}_2({\bf 0})$ is the pair
correlation function on length scales that are greater than the range of the
interaction but small compared with other lengths in the problem, such as
the thermal de Broglie wavelength, the particle separation, and, when a
condensate is present, the coherence length.  We have chosen the normalization
of the wave function such that at distances large compared with the range of the
$1s-1s$ potential it behaves as $\Psi_{\rm rel}(r) \simeq 1-a_{1s-1s}/r$. Note
that the above procedure is equivalent to assuming a wave function of the
Jastrow form to describe the correlations at short distances, the Jastrow factor
being taken to  be of the form of the relative wave function of two atoms at
zero energy \cite{jastrow}.

The final result for the shift is thus
\begin{equation}
\overline{\Delta \omega} = \frac{n}{\hbar} g^{\rm MF}_2({\bf 0}) \int d{\bf r}~
[V_{1s-2s}(r)-V_{1s-1s}(r)] |\Psi_{\rm rel}(r)|^2~.
\end{equation}
This expression cannot be simply rewritten in terms of scattering lengths.
A simple example that demonstrates this is a $1s-1s$ interaction with a hard
core at a radius $r_c$. The expression for the frequency shift does not
depend on the $1s-2s$ potential at distances less than $r_c$, since the relative
wave function for two atoms in the $|1s\rangle$ state vanishes there. However,
the $1s-2s$ scattering length is sensitive to the behavior of the $1s-2s$
potential at distances less than $r_c$, and therefore this is incompatible with
the frequency shift being expressible solely in terms of scattering lengths. We
expect the Jastrow form of the wave function to be accurate irrespective of
whether or not the gas is Bose condensed, and consequently in a completely
Bose-condensed gas the shifts are predicted to be a factor of 2 smaller than in
a gas of the same density with no condensate, reflecting the usual $2!$
reduction factor for two-body processes \cite{henk}.

The result of this calculation is that the frequency-weighted sum rule is
quite different from what one predicts if one uses the pseudopotential.  To
understand the origin of these differences it is convenient to explore the
problem from a microscopic viewpoint.

\section{Microscopic approach}
\label{ma}
To understand the sum-rule result, it is helpful to think about the
nature of the final states that can be created from the initial state
by the operator of interest, which in this case converts a $1s$ atom
into a $2s$ one. Relative to the initial state, the simplest excited
states have an extra $2s$ quasiparticle and an extra $1s$ quasihole, and
will be referred to as single quasiparticle-quasihole pair excitations. In the
random phase approximation these are the only states take into
account. The physics of the process may be understood by regarding
the degree of freedom associated with converting a $1s$ atom
into a $2s$ one as a pseudospin.  If the commutator of the
pseudospin-raising operator with the unperturbed Hamiltonian is zero,
there is a unique frequency for all transitions.
Because the interaction between a $1s$ atom and a $2s$ atom
differs from that between two $1s$ atoms, however, the Hamiltonian is not
invariant under rotations in pseudospin space, and its commutator with
the pseudospin-raising operator is not zero. Consequently there
can be transitions to states with a range of energies. It is perhaps
helpful to consider a spin system in an applied magnetic field. If
the interaction between the particles commutes with the spin-raising
operator, the raising operator will couple only to states whose energy differs
from that of the original state by $\hbar$ times the Larmor frequency.
However, if the interaction is not invariant under spin rotations,
other excited states with different energies can be created. In
Fermi-liquid theory the first sort of transitions correspond to the
creation of a single quasiparticle-quasihole pair, while the more complicated
excitations correspond to creation of many pairs. For the problem
under study here, an $n$-pair excitation has one extra $2s$
quasiparticle, $n$ extra $1s$ quasiparticles, and $n+1$ extra $1s$
quasiholes. The difference between the results for the
frequency-weighted sum rule calculated with the pseudopotential and
the actual potential is due to the multipair excitations. For Fermi
liquids an analysis of the density response in terms of single-pair
and multipair states may be found in Ref.~\cite{pn}. A formulation of the
problem for more general sorts of response was presented in terms of
microscopic theory by Leggett \cite{leggett}, and the results were discussed
in terms of Fermi-liquid theory in Ref.~\cite{bp}.

\begin{figure}
\psfig{figure=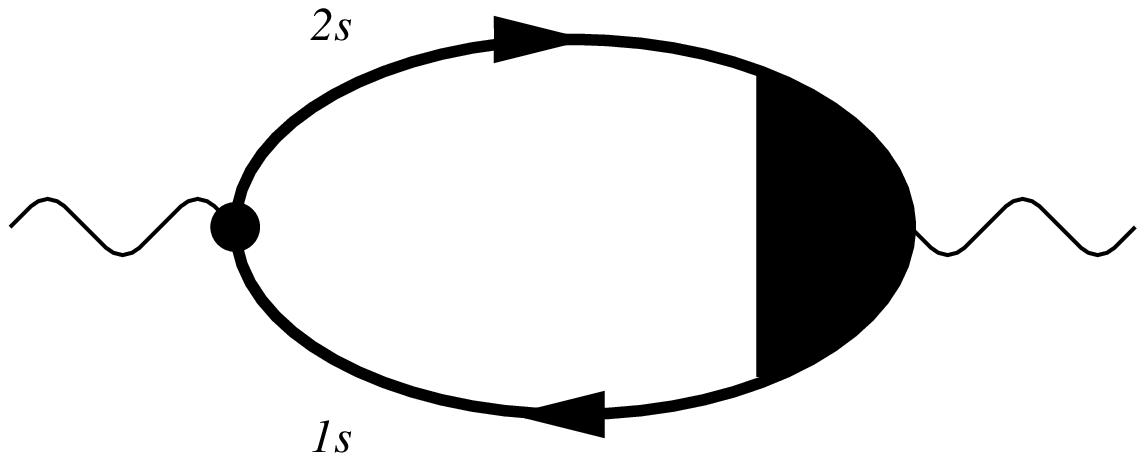}
\caption{The polarization diagram that determines the two-photon absorption
         line shape. The thick lines denote the exact $1s$ and $2s$
         propagators, and the small and large black areas denotes the
         bare and exact vertex functions, respectively.
         \label{pol}}
\end{figure}

Let us begin by expressing the result for the transition rate in terms of the
response function for the operator
\begin{equation}
{\cal O} = \frac{1}{V} \int d{\bf x}
  \left( e^{-i\omega t}\psi_{2s}^{\dagger}({\bf x})\psi_{1s}({\bf x}) +
         e^{i\omega t}\psi_{1s}^{\dagger}({\bf x})\psi_{2s}({\bf x}) \right)~,
\end{equation}
where $V$ is the volume of the system.
The response function is defined in the usual way as the temporal Fourier
transform of the retarded commutator, and is given by
\begin{eqnarray}
\label{self}
\chi(\omega)&=& - \frac{1}{V}\int d{\bf x} \int_0^{\infty} dt~ e^{i\omega t}
       \langle [\psi_{1s}^{\dagger}({\bf x},t) \psi_{2s}({\bf x},t),
                 \psi_{2s}^{\dagger}({\bf 0},0) \psi_{1s}({\bf 0},0)]\rangle
\nonumber \\
&=&\sum_{m,n} \frac{|\langle m|{\cal
O}|n\rangle|^2 p_n}{\hbar\omega +i0 + E_n  - E_m}~,
\label{chi}
\end{eqnarray}
where $p_n$ is again the probability of the state $n$ being occupied and
we neglected the occupancy of the final state compared with that of the
initial one. The transition rate in Eq.~(\ref{gr}) is therefore given by
\begin{equation}
I(\omega) = - \frac{2}{\hbar} \Im \left[ \Pi(\omega) \right],
\end{equation}
where
\begin{equation}
\label{pola}
\Pi(\omega) =   \left(\frac{\hbar\Omega}{2}\right)^2 \chi(\omega)
\end{equation}
is the polarizability of the gas.  This is the
desired result, because it explicitly shows that the transition rate is
related to the polarizability of the gas, which is easily accessible with
equilibrium many-body techniques. Indeed, in that language
$\hbar\Pi(\omega)$ is equal to the (retarded) self energy for
the ``effective photon'' causing the $1s-2s$ transition and the imaginary part
therefore determines its finite lifetime, which physically is due to absorption
by $1s$ atoms in the gas. We are thus left with the task
of calculating the polarizability, which theoretically implies that we have to
evaluate the diagram in Fig.~\ref{pol}. We begin by considering two simple
calculations, the  Hartree-Fock approximation and the random phase
approximation, before discussing the more general formulation.

\subsection{The Hartree-Fock and Random Phase Approximations}
\label{mf}
In this section we consider a number of examples where only coherent
contributions to the response are taken into account. These calculations
lead to results identical with those of Oktel and Levitov \cite{ol}. To
familiarize ourselves with the present formulation, let us first consider the
ideal Bose gas. Then Eq.~(\ref{self}) becomes
\begin{equation}
\chi(\omega)
= - \frac{1}{V} \int d{\bf x} \int_0^{\infty} dt~ e^{i\omega t}
        \langle \psi_{1s}^{\dagger}({\bf x},t) \psi_{1s}({\bf 0},0) \rangle
       \langle \psi_{2s}({\bf x},t) \psi_{2s}^{\dagger}({\bf 0},0) \rangle~.
\end{equation}
Moreover, the single-particle propagator is given by
\begin{equation}
\langle \psi_{1s}^{\dagger}({\bf x},t) \psi_{1s}({\bf 0},0) \rangle
  = \frac{1}{V} \sum_{\bf k}~ N_{\bf k}
      e^{-i {\bf k} \cdot {\bf x} +
          i (\epsilon_{\bf k} + \epsilon_{1s} - \mu)t/\hbar} ~,
\end{equation}
where $\epsilon_{\bf k} = \hbar^2{\bf k}^2/2m$ is the kinetic
energy of a $1s$ atom, $\mu$ is the chemical potential for 1$s$
atoms, and $N_{\bf k} = 1/(e^{\beta(\epsilon_{\bf k}-\mu)}-1)$ is
the Bose distribution function with $\beta = 1/k_BT$. Similarly we
have, including now the finite atomic lifetime of the $2s$ atom,
\begin{equation}
\langle \psi_{2s}({\bf x},t) \psi_{2s}^{\dagger}({\bf 0},0) \rangle
  = \frac{1}{V} \sum_{\bf k}
      e^{i {\bf k} \cdot {\bf x} -
         i (\epsilon_{\bf k} + \epsilon_{2s} - i\hbar\Gamma_{2s}/2 -
                                                            \mu)t/\hbar}~.
\end{equation}
Substituting the last two results, we find for the polarizability in
Eq.~(\ref{pola}) the expression
\begin{eqnarray}
\Pi(\omega) &=& - i \frac{nV\hbar\Omega^2}{8}
    \int_0^{\infty} dt~ e^{i\omega t}
                e^{i(\epsilon_{1s}-\epsilon_{2s}+i\hbar\Gamma_{2s}/2)t/\hbar}
                                                             \nonumber \\
   &=& \frac{nV(\hbar\Omega)^2}{8}
          \frac{1}{\hbar\omega + \epsilon_{1s} - \epsilon_{2s} +
                                                      i\hbar\Gamma_{2s}/2}~.
\end{eqnarray}
Therefore, we conclude that the absorption line of the gas has a profile given
by
\begin{equation}
\label{al}
I(\omega) = N \frac{(\hbar\Omega)^2}{8}
              \frac{\Gamma_{2s}}
                   {(\hbar\omega - (\epsilon_{2s}-\epsilon_{1s}))^2 +
                                                    (\hbar\Gamma_{2s}/2)^2}~,
\end{equation}
which is just the number of atoms $N = nV$ times the atomic line profile and
exactly centered at the
atomic resonance in this case. Note that diagrammatically we have now
calculated the lowest order contribution to the polarizability in
Fig.~\ref{pol}, in which the
exact $1s$ and $2s$ propagators are replaced by the ideal gas
ones and there are no vertex corrections.

\begin{figure}
\psfig{figure=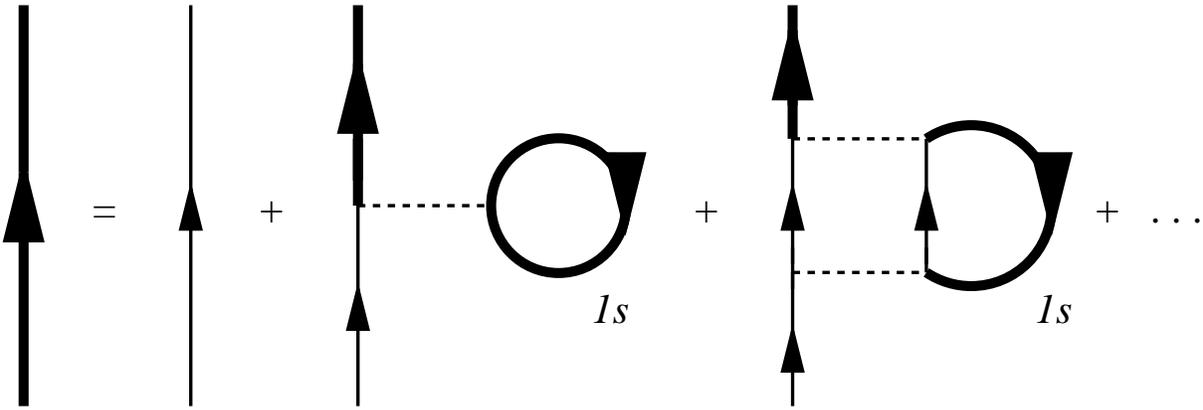}
\caption{The $1s$ and $2s$ propagators in
         the T-matrix approximation. The
         thin lines represent the bare propagators and the dashed lines the
         interactions.
         \label{prop}}
\end{figure}

At the next level of approximation we dress the $1s$ and $2s$ propagators by
including the effect of atom-atom ladder diagrams as shown in Fig~\ref{prop}.
This corresponds to
a Hartree-Fock approximation, in which the effective interaction is taken
to be the T matrix for two-body scattering.
For a gas with no condensate, the effect of  dressing the propagators in the
above calculation is to  replace $\epsilon_{\alpha}$ by
$\epsilon_{\alpha} + \hbar\Sigma_{\alpha}$, where to lowest order in the
T matrix
\begin{equation}
\label{hf1s}
\hbar\Sigma_{1s} = \frac{8\pi a_{1s-1s}
\hbar^2 n}{m}~,
\end{equation}
and
\begin{equation}
\label{hf2s}
\hbar\Sigma_{2s} = \frac{4\pi a_{1s-2s} \hbar^2 n}{m}~.
\end{equation}
The factor-of-two difference between the numerical factors in Eqs.~(\ref{hf1s})
and (\ref{hf2s}) reflects the fact that both the
Hartree and Fock terms contribute to the energy of a $1s$ atom, but only
the Hartree term contributes for a pair of unlike atoms. Because these
interaction corrections to the atomic energies are purely real,  the
absorption line is of the same shape as in Eq.~(\ref{al}), but is now
centered at a frequency shifted from the single-atom resonance by an
amount
\begin{equation}
(\Delta\omega)_{\rm HF} = \frac{4\pi\hbar n}{m} (a_{1s-2s} - 2a_{1s-1s})~.
\end{equation}
This is the ``naive'' Hartree-Fock result for the
collisional frequency shift due to the mean-field interaction that a $1s$ and
a $2s$ atom experience from the surrounding gas of $1s$ atoms. Most important
for our purposes is that if we repeat the above calculation for a fully Bose
condensed gas of $1s$ atoms, we find that now
\begin{equation}
(\Delta\omega)_{\rm HF} = \frac{4\pi\hbar n}{m} (a_{1s-2s} - a_{1s-1s})~,
\end{equation}
which shows that only the contribution from the $1s-1s$ mean-field interaction
is reduced by a factor of $2$, in agreement with the fact that $1s$ and $2s$
atoms are distinguishable.

\begin{figure}
\psfig{figure=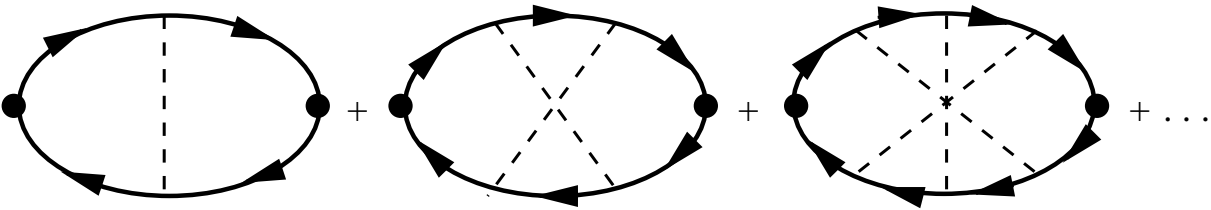}
\caption{The vertex corrections in the T-matrix approximation.
         \label{vert}}
\end{figure}

We have called the above Hartree-Fock approximation naive, because it is well
known that for an approximation to satisfy the conservation laws, it is
necessary to include vertex corrections in addition to the self energy
corrections discussed above. For this problem this amounts to including the
effects of the mean field self-consistently.  Formally, these two kinds of
corrections are related by the condition that the vertex correction must be
the functional derivative of the self energy corrections with respect to the
applied field \cite{baym}. In our case this implies that we also have to
calculate the  ``maximally crossed'' diagrams shown in Fig.~\ref{vert}. These
correspond to the chains of particle-hole bubble diagrams calculated in
the random phase approximation and in Fermi liquid theory. This is easily
achieved since it corresponds to summing the geometric series
\begin{eqnarray}
\frac{1}{\hbar\omega - (\Delta\omega)_{\rm HF} + \epsilon_{1s} - \epsilon_{2s}
                                                  + i\hbar\Gamma_{2s}/2} +
         \hspace*{3.7in} \nonumber \\
\frac{1}{\hbar\omega -
         (\Delta\omega)_{\rm HF} + \epsilon_{1s} - \epsilon_{2s}
                           + i\hbar\Gamma_{2s}/2}   \frac{4\pi
a_{1s-2s}\hbar^2 n}{m}  \frac{1}{\hbar\omega - (\Delta\omega)_{\rm HF} +
\epsilon_{1s} - \epsilon_{2s}
     + i\hbar\Gamma_{2s}/2} + \dots~.                       \nonumber
\end{eqnarray}
In the end we thus find that
\begin{equation}
I(\omega) = N \frac{(\hbar\Omega)^2}{8}
   \frac{\Gamma_{2s}}
        {(\hbar\omega - (\Delta\omega)_{\rm RPA}
                      - (\epsilon_{2s}-\epsilon_{1s}))^2 +
                                                    (\hbar\Gamma_{2s}/2)^2}~,
\end{equation}
with
\begin{equation}
(\Delta\omega)_{\rm RPA}
             = (\Delta\omega)_{\rm HF} + \frac{4\pi a_{1s-2s}\hbar n}{m}
             = \frac{8\pi\hbar n}{m} (a_{1s-2s} - a_{1s-1s})~.
\end{equation}
Moreover, in the fully Bose-Einstein condensed case the vertex corrections are 
absent and
we recover the Hartree-Fock result,
\begin{equation}
(\Delta\omega)_{\rm RPA} = (\Delta\omega)_{\rm HF} =
                             \frac{4\pi\hbar n}{m} (a_{1s-2s} - a_{1s-1s})~.
\end{equation}
We therefore conclude that for a fully condensed Bose gas the collisional
frequency shift is indeed reduced by an overall factor of 2, in agreement
with our sum-rule result in Sec.~\ref{sr} and the work of Ref.\ \cite{ol}.

\subsection{Coherent and Incoherent Contributions to Response Functions}
\label{cinc}
To explore the physics in greater detail, it is convenient to adopt an
approach exploited in the context of Fermi-liquid theory \cite{leggett}.
We first expresses the single-particle propagator in a
many-body system as the sum of a coherent part coming from an
intermediate state with a
single quasiparticle excitation, and an incoherent part
coming from more complicated excitations. Mathematically this implies that
\begin{equation}
G^{(2)}({\bf p}, \epsilon)=  G^{(2)}_{\rm coh}({\bf p}, \epsilon)
+G^{(2)}_{\rm inc}({\bf p}, \epsilon)~,
\end{equation}
where $\bf p$ is the momentum, and $\epsilon$ is the energy.
The coherent part, which corresponds to the quasiparticle, is given by
\begin{equation}
G^{(2)}_{\rm coh}({\bf p}, \epsilon)
  =\frac{Z({\bf p})}{\epsilon-\epsilon({\bf p})}~,
\end{equation}
where $Z({\bf p})$ is the renormalization factor or quasiparticle residue, and
$\epsilon({\bf p})$ is the
quasiparticle energy. Both of these quantities
depend on the atomic species considered.  The incoherent contribution to the
propagator corresponds to transient effects due to the dressing of a
free atom to make it into a quasiparticle. Likewise the two-particle propagator
$G^{(4)}({\bf p}, \epsilon; {\bf p}', \epsilon')$ for a pseudospin fluctuation
may
be expressed in terms of a coherent part, corresponding to a single
quasiparticle-quasihole pair, plus an incoherent part coming from
multipair excitations, i.e.,
\begin{equation}
G^{(4)}({\bf p}, \epsilon; {\bf p}', \epsilon') =
 \frac{Z_{1s}({\bf p})}{(\epsilon-\epsilon_{1s}({\bf p}))}
 \frac{Z_{2s}({\bf p}')}
      {(\epsilon'-\epsilon_{2s}({\bf p}'))} +
G^{(4)}_{\rm inc}({\bf p}, \epsilon;{\bf p}', \epsilon')~.
\end{equation}

We next analyse the diagrams for the response function $\chi$
by separating the single particle
propagators into their coherent and incoherent contributions, as was
done by Leggett in the context of Fermi systems. We then divide these
diagrams into two classes. The first class contains those diagrams which are
reducible with respect
to the coherent contributions of two (one $1s$ and one $2s$) single-particle
propagators.
We refer to these as the coherent contribution. The second class contains
all diagrams which are not reducible in this sense and we call this the
incoherent contribution. We remark that since the operator $\cal O$ does
not change the total particle number, the two coherent particle lines
must have their arrows in opposite directions, and therefore
correspond to a quasiparticle-quasihole pair.

Expressed in a
formal matrix notation the response function may be written as
\begin{equation}
\chi ={\rm Tr[}G^{(4)}(1-\Gamma^{(4)} G^{(4)})^{-1}]~,
\end{equation}
where $\Gamma^{(4)}$ is the two-particle vertex function that is irreducible
with respect to two particle lines with oppositely directed arrows,
i.e., it is irreducible in the particle-hole channel. Separating out
the terms that contain only incoherent contributions to $G^{(4)}$
from the others, we find
\begin{equation}
\chi =\chi_{\rm inc} +\chi_{\rm coh}~,
\end{equation}
where
\begin{equation}
\chi_{\rm inc} = {\rm Tr}[G^{(4)}_{\rm inc}(1-\Gamma^{(4)}
                          G^{(4)}_{\rm inc})^{-1}]
\end{equation}
and
\begin{equation}
\chi_{\rm coh} =
{\rm Tr}[(1-\Gamma^{(4)} G^{(4)}_{\rm inc})^{-1}G^{(4)}_{\rm coh}
(1-\Gamma^{(4)}_{\rm coh}G^{(4)}_{\rm coh})^{-1}
(1-\Gamma^{(4)} G^{(4)}_{\rm inc})^{-1}]~.
\end{equation}
The factor $(1-\Gamma^{(4)} G^{(4)}_{\rm inc})^{-1}$ corresponds to a vertex
renormalization and the quantity
\begin{equation}
{\Gamma}_{\rm coh}^{(4)}=\Gamma^{(4)} (1-G^{(4)}_{\rm inc}\Gamma^{(4)})^{-1}
\end{equation}
is a renormalization of the interactions between the coherent parts of a
particle-hole excitations due to intermediate states with incoherent
particle-hole pairs.
For the present problem, an important feature of this result is
the existence of the incoherent contribution to $\chi$, since this is
what is responsible for the difference between the sum rule evaluated
with the pseudopotential and the true sum rule.

We turn now to the coherent contribution to the response function.
The coherent part of $G^{(4)}$ has the same form as for two particles
with energies modified by the medium, apart from the renormalization
factors $Z$.  However, if one multiplies the matrix element
for coupling of the two photons to the excitations by a
factor $ (1-\Gamma^{(4)} G^{(4)}_{\rm inc})^{-1}(Z_{2s} Z_{1s})^{1/2}$
and uses for the  effective
interaction between a quasiparticle and a quasihole the quantity
$Z_{1s}Z_{2s} {\Gamma}^{(4)}$, the coherent contribution to the
response has precisely the same form as in the random phase approximation
calculation above. This modified interaction plays a role analogous
to that of the quasiparticle-quasiparticle interaction introduced in
Fermi-liquid theory by Landau.

Let us now analyse the consequences of the above for a low density gas.
In that case the renormalization
factors tend to unity, and the two-particle vertex reduces to the T
matrix. The quasiparticle energies reduce to the Hartree-Fock ones,
and the mean-field interaction is also just the T matrix. Thus the
coherent contribution to the response has precisely the form predicted
by the mean-field theory calculation in Sec.~\ref{mf}. Observe that in 
calculating the average frequency associated with the coherent part of the 
response, the renormalization factor for the effective two-photon matrix element 
cancels out.  What implications does
our calculation have for experiment?
In addition to a sharp peak in the absorption due to the excitation of
a single quasiparticle-quasihole pair, the calculation predicts a broad 
background
due to creation of more complicated final states. However, because
the background is expected to be a rather smoothly-varying function of
frequency that extends over a large frequency range, it is difficult to 
detect. Consequently, the part of
the absorption spectrum that is investigated experimentally is only
that due to the coherent contribution to the response function.

\section{Collisional broadening}
\label{cb}
We now consider how collisions broaden the coherent part of the line. One
effect is that the self energy of the atoms acquires an imaginary part, and
the coherent parts of the propagators become
\begin{equation}
G^{(2)}_{\rm coh}({\bf p}, \epsilon)
 =\frac{Z({\bf p})}{\epsilon-\epsilon({\bf p})
            +i\hbar\Sigma''({\bf p}, \epsilon({\bf p}))}~,
\end{equation}
where $\Sigma''({\bf  p}, \epsilon)$ is the imaginary part of the
self energy. Another effect is that there are vertex corrections
analogous to those responsible for the contributions to the line
shift beyond what is predicted by the Hartree-Fock approximation.
The total width is most easily calculated by observing that the
propagation of a $2s$ atom and a $1s$ hole is determined by the
difference between the $1s-2s$ interaction and the $1s-1s$
interaction. For definiteness, let us consider a gas with no
condensate. In the absence of the $1s-2s$ interaction, the only
contribution to the width comes from the imaginary part of the
self energy of the $1s$ atom, which is given in the dilute limit
by \cite{stoof}
\begin{eqnarray}
\hbar\Sigma''_{1s}({\bf p},\epsilon_{\bf p})
   &=& - 2\pi \left(\frac{4\pi a_{1s-1s}\hbar^2}{m}\right)^2
   \frac{1}{V^2}
   \sum_{{\bf p'} {\bf p''}}~ \delta(\epsilon_{\bf p}+\epsilon_{\bf p'}
                      -\epsilon_{\bf p+p''}-\epsilon_{\bf p'-p''}) \nonumber \\
     &\times& [N_{\bf p'}(1+N_{\bf p'-p''})(1+N_{\bf p+p''})
                           - (1+N_{\bf p'})N_{\bf p'-p''}N_{\bf p+p''}]~,
\end{eqnarray}
the factor of two being the result of the Bose enhancement of the
cross section, which is due to the exchange process. Since the
imaginary part of the self energy is momentum dependent the
absorption line of the gas is in principle not exactly Lorentzian.
Nevertheless, the typical width of the line is determined by the
average $-(2/N)\sum_{\bf p} N_{\bf p} \Sigma''_{1s}({\bf
p},\epsilon_{\bf p})$ and thus equals
\begin{eqnarray}
\Delta\Gamma_{2s}
   &=& \frac{(4\pi\hbar)^3 (a_{1s-1s})^2}{nm^2}
     \frac{1}{V^3} \sum_{{\bf p} {\bf p'} {\bf p''}}
       \delta(\epsilon_{\bf p}+\epsilon_{\bf p'}
               -\epsilon_{\bf p+p''}-\epsilon_{\bf p'-p''})
               \nonumber \\
   &\times &  N_{\bf p}[N_{\bf p'}(1+N_{\bf p'-p''}) (1+N_{\bf p+p''})-
             (1+N_{\bf p'})N_{\bf p'-p''}N_{\bf p+p''}]~.
\end{eqnarray}
When the $1s-2s$ interaction is included, the result is simply
\begin{eqnarray}
\Delta\Gamma_{2s}
   &=& \frac{(4\pi\hbar)^3 (a_{1s-1s}-a_{1s-2s})^2}{nm^2}
     \frac{1}{V^3} \sum_{{\bf p} {\bf p'} {\bf p''}}
       \delta(\epsilon_{\bf p}+\epsilon_{\bf p'}
               -\epsilon_{\bf p+p''}-\epsilon_{\bf p'-p''})
               \nonumber \\
   &\times &  N_{\bf p}[N_{\bf p'}(1+N_{\bf p'-p''}) (1+N_{\bf p+p''})-
             (1+N_{\bf p'})N_{\bf p'-p''}N_{\bf p+p''}]~.
\end{eqnarray}
In the classical limit, this reduces to
\begin{equation}
\Delta\Gamma_{2s} =  8\pi n (a_{1s-1s}-a_{1s-2s})^2
                       \langle v_{\rm rel} \rangle~,
\end{equation}
where $\langle v_{\rm rel} \rangle=4 (kT/\pi m)^{1/2}$ is
the average relative velocity between two $1s$ atoms in the atomic
hydrogen gas. The total width of the line is the sum of the
natural width and the collisional contribution, and it is
therefore equal to $\Gamma_{2s}+\Delta\Gamma_{2s}$. The width is
of order $(a_{1s-2s}-a_{1s-1s})/\lambda_T$ times the shift, where
$\lambda_T=\hbar/(2\pi m k_B T)^{1/2}$ is the thermal de Broglie
wavelength. When a condensate is present, the above calculation
can be easily generalized. In a first approximation we only need
to take into account explicitly the macroscopic occupation of the
zero-momentum state by substituting $N_{\bf p} \rightarrow n_c V
\delta_{{\bf p},{\bf 0}} + N_{\bf p}$, with $n_c$ the condensate
density. At the next level of approximation we also need to
incorporate the Bogoliubov coherence factors.

\section{CONCLUSIONS}
\label{concl}
In this paper we have considered the effect of interactions on the two-photon
absorption line profile in spin-polarized atomic hydrogen by means of a
frequency-weighted sum rule and by means of microscopic many-body theory. We
have shown that the line profile consists of a narrow coherent peak on top of a
broad incoherent background.  For typical atomic potentials this background in 
principle has sufficient
spectral weight that the pseudopotential approximation does not give an accurate 
estimate of the total contribution to the
frequency-weighted sum rule. However, the frequency of the narrow peak, which is 
the feature most easily seen
experimentaly, may be expressed in terms of the
low-energy pseudopotentials.

We have also shown that the collisional frequency shift of the
absorption line is reduced by a factor of two if the gas is fully
Bose condensed. We have pointed out that for this factor-of-two
reduction it is crucial to take many-body correlation effects into
account that go beyond the Hartree-Fock approximation 
commonly used for these dilute atomic gases. At this point it is
worth mentioning that the Bose-Einstein condensation experiments
by Fried {\it et al.} apparently do not to see this effect. Their results seem 
to be consistent with 
the Hartree-Fock theory, which, due to
the fact that $a_{1s-1s} \ll |a_{1s-2s}|$ for atomic hydrogen,
basically predicts no reduction at all \cite{mit1}. At present we
have no explanation for the cause of this discrepancy.

\section*{ACKNOWLEDGMENTS}
We are grateful to our experimental colleagues T.\ Greytak, T.\ C.\ Killian, D.\ 
Kleppner, and
L.\ Willmann for helpful discussions and correspondence.  In addition, we
greatly appreciate correspondence with L.\ Levitov, who kindly provided us with a
draft of reference \cite{levitov}. We also thank A.\
Dalgarno and M.\ J.\ Jamieson for helpful communications with respect to the
interaction potentials. One of us (CJP) is grateful to the participants at the
Aspen Institute for Physics workshop on Bose-Einstein condensation for useful
discussions.  Part of this work was carried out while we participated in the 
Lorentz Center workshop on Bose-Einstein condensation.

\end{document}